\documentclass[12pt]{article}
\usepackage[top=0.9in,bottom=0.9in,left=0.8in,right=0.8in]{geometry}
\usepackage{rotating}
\usepackage{caption}
\usepackage{amsmath}
\usepackage{amsfonts}
\usepackage{lscape}
\usepackage{float}
\usepackage{booktabs}
\usepackage{url}
\usepackage{threeparttable}
\usepackage{multirow}
\usepackage{color}
\usepackage{amsmath}
\usepackage{graphicx}
\usepackage{bm}
\usepackage{lipsum}
\usepackage{makecell}
\usepackage[sectionbib,round]{natbib}
\usepackage{pdfpages}
\usepackage{multirow}
\usepackage{authblk}
\usepackage{booktabs} 
\usepackage{pdfpages}

\usepackage{extarrows,amsmath,amssymb,comment}
\usepackage[FIGTOPCAP]{subfigure}
\usepackage{subfig}
\usepackage{subfigure}

\usepackage{amssymb}

\usepackage{amsthm}

\begin{document}
\title{\bf Allocation of COVID-19 Testing Budget on a Commute Network of Counties 
}

\author[1]{\small Yaxuan Huang}
\author[2]{\small Zheng Tracy Ke*}
\author[3]{\small Jiashun Jin}

\affil[1]{\footnotesize Department of Statistics, University of California, Berkeley, California, USA}
\affil[2]{\footnotesize Department of Statistics, Harvard University, Massachusetts, USA}
\affil[3]{\footnotesize Department of Statistics, Carnegie Mellon University, Pennsylvania, USA}

\date{}


\maketitle

\noindent
{\bf Abstract: }{The screening testing is an effective tool to control the early spread of an infectious disease such as COVID-19. When the total testing capacity is limited, we aim to optimally allocate testing resources among $n$ counties. We build a (weighted) commute network on counties, with the weight between two counties a decreasing function of their traffic distance. We introduce a network-based disease model, in which the number of newly confirmed cases of each county depends on the numbers of hidden cases of all counties on the network. Our proposed testing allocation strategy first uses historical data to learn model parameters and then decides the testing rates for all counties by solving an optimization problem. We apply the method on the commute networks of Massachusetts, USA and Hubei, China and observe its advantages over testing allocation strategies that ignore the network structure. Our approach can also be extended to study the vaccine allocation problem.}


\maketitle

\section{Introduction}\label{sec:Intro}

Since December 2019, the severe acute respiratory syndrome coronavirus 2 (SARS-Cov-2) has had a rapid spread over the world. By September 30, 2021, there have been 0.23 billion confirmed cases of COVID-19, including 4.8 million deaths. Many studies show the necessity of interventions in the early stage of the pandemic \citep{hao2020reconstruction,wang2020epidemiological}. The screening testing is an effective intervention approach. COVID-19 has the characteristics of high infectiousness, a considerable proportion of asymptomatic infections, a long incubation period, and early symptoms hard to distinguish from other diseases such as influenza. Therefore, the active screening of hidden cases, followed by subsequent quarantines and treatments, plays a key role in blocking the virus transmission path more timely and thoroughly, especially before vaccines are available. 

A common testing method for COVID-19 is the SARS-Cov-2 nucleic acid test. The daily testing capacity is usually limited, because of the limit on producing test kits and the manpower on sample collection and lab work. We are interested in developing optimal strategies to allocate the limited testing resources. Existing strategies include utilization of the population age structure, priority to healthcare workers, contact tracing (e.g., testing of the close contacts of confirmed cases), and group testing to increase efficiency. In this paper, we consider a scenario where a state has a limited daily testing budget and aims to optimally allocate it to different counties. We propose to design the allocation strategy based on a `commute network' of counties and the history of confirmed cases. The commute network is a weighted network, where the weighted edge between two counties is a decreasing function of the traffic distance between them. A high-degree node (county) in the network has small traffic distances with many other counties, so that a high infection rate of this county is more likely to cause a rapid spread of disease in the state.  
Our allocation strategy will give priority to those counties that have a high degree in the commute network or a large number of confirmed cases in the past. 

We introduce a 4-compartment SIR disease model on the network. The four compartments are S (susceptible), H (hidden), C (confirmed) and R (removed), where individuals in compartments H and C are both infected and infectious. 
We choose these four compartments by taking into account special characteristics of COVID-19: (i) The infected individuals soon become infectious, hence, we do not model the exposure stage. (ii) There are a considerable fraction of asymptomatic cases \citep{Byambasuren2020}, making the timely diagnosis not easy; hence, we divide the infected individuals into two compartments, H and C. 
We note that U.S. and many other countries have had strict isolation policies for confirmed cases. For this reason, we make an assumption that the main resource of infections comes from those undiagnosed infected cases, i.e., the individuals in compartment H. The goal of screening testing is to minimize the number of individuals in compartment H.  
Suppose a state has $n$ counties. Let $(S_i(t), H_i(t), C_i(t), R_i(i))$ denote the respective number of individuals in four compartments at county $i$, for $1\leq i\leq n$. We propose an ordinary difference equation model for $\{(S_i(t), H_i(t), C_i(t), R_i(t))\}_{1\leq i\leq n}$, where the number of newly infected cases of each county depends on the number of past hidden cases of neighbor counties on the network (including the county itself), as well as the allocated testing rate of this county. 
The state administrator has to decide the allocation rates of all counties to minimize the total number of hidden cases in the future. 
We solve this problem by a 2-stage procedure. In the first stage, we use data of newly confirmed cases of all counties at $t=1,2,\ldots,t_0-1$ to estimate model parameters. 
In the second stage, we solve an optimization to minimize $\sum_{t=t_0}^{T}\sum_{i=1}^n H_i(t)$.  

We implement our method on the commute network for Massachusetts, United States and the commute network for Hubei, China (the traffic distances are from online map services, and the disease data are simulated from our model). We observe an advantage of our method over the allocation strategies that ignore network structures, e.g., allocation by county populations and allocation by county-wise infection rates.

There have been studies on the cross-region allocation of testing resources for an infectious disease. \cite{Yin2021} proposed to optimize the distribution of treatment centers and resources among regions to minimize the total number of new infections and funerals. \cite{Baunez2020} studied the sub-national allocation of testing resources in Italy, using the slopes of confirmed cases versus test numbers in different regions. \cite{buhat2020optimal} proposed to use natural language processing tools to allocate test kits more fairly among test centers in Philippines. However, none of these methods are based on the commute network structure. 

\cite{ou2020and} proposed a framework for active screening of a SIS disease based on a contact tracing network. In their model, the disease has two compartments, S (susceptible) and I (infected). Each node in the network is an individual, and the state variable of a node is a binary variable indicating whether this individual is infected. They introduced a model to characterize how the network structure and screening strategy affect disease spread. They also proposed an optimization algorithm to compute the optimal testing allocation strategy. 
Our disease model and optimization algorithm borrow ideas from \cite{ou2020and}, but there are major differences. We consider a SIR disease with four compartments, S, H, C and R. Our network is a commute network, where each node is a county and the state variable of a node is the 4-dimensional vector containing the number of individuals in each compartment. Our model, parameter estimation, and computation of allocation strategies are different from those in \cite{ou2020and}.  

The remainder of this paper is organized as follows: Section~\ref{sec:Methods} contains our main results, including the disease model, parameter estimation, and computation of the optimal allocation strategy. Section~\ref{sec:RealData} contains semi-synthetic experiments on the real commute networks for Massachusetts, USA and Hubei, China. 
Section~\ref{sec:Vaccine} generalizes our model and algorithm to the problem of vaccine allocation. Section~\ref{sec:Discuss} concludes the paper.

\section{The network-based active screening strategy}\label{sec:Methods}
\begin{figure}[tb!]
\centering
\includegraphics[width=7cm]{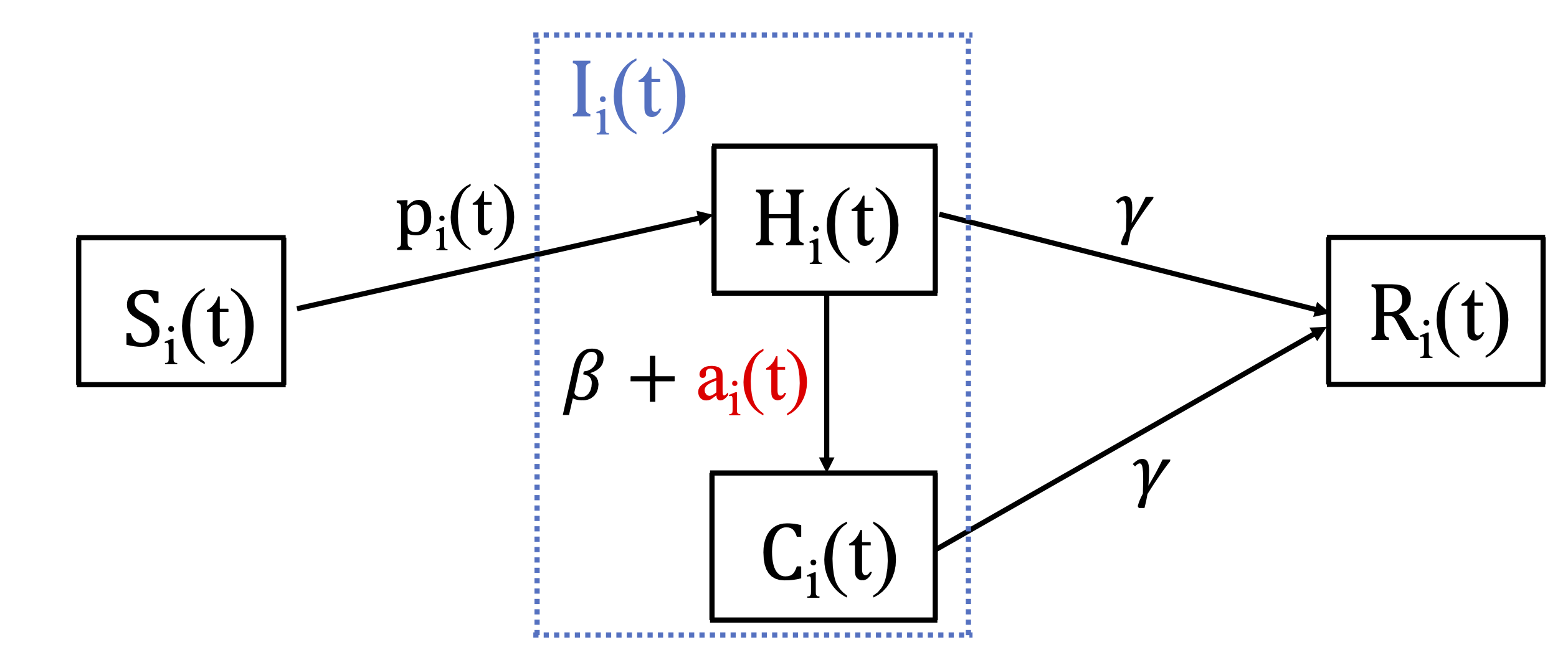}
\caption{The disease model. Here, $S_i(t)$, $H_i(t)$, $C_i(t)$ and $R_i(t)$ are the respective number of susceptible, hidden, confirmed and removed individuals at county $i$, $\beta$ is the (baseline) diagnosis rate, $\gamma$ is the recovery rate, $a_i(t)$ is the allocated testing rate of county $i$, and $p_i(t)$ is the infection probability of county $i$. The infection probability $p_i(t)$ depends on the number of hidden cases of other counties through the commute network.}\label{fig:compartments}
\end{figure}

In a SIR disease model, the population is divided into three compartments: S (susceptible), I (infected, including the asymptomatic, presymptomatic and symptomatic cases), and R (removed, including the dead, recovered and vaccinated cases). Since an infected individual may not be diagnosed, we further divide compartment I into two sub-compartments: H (hidden) and C (confirmed). This gives rise to four compartments: S, H, C and R, as shown in Figure~\ref{fig:compartments}.  
Suppose there are $n$ different counties. For simplicity, we assume the population of every county is time-invariant during the period of interest. Let $N_i$ denote the population of county $i$, for $1\leq i\leq n$, and write $\mathbf{N}=(N_1, N_2,\ldots,N_n)^{\top}$. Let $S_i(t)$, $H_i(t)$, $C_i(t)$ and $R_i(t)$ be the respective number of individuals in four compartments for county $i$ at time $t$. 
Furthermore, let $C_i^{new}(t)$ be the number of newly confirmed cases, where a newly confirmed case is an individual transferred from compartment H to compartment C.  Write $\mathbf{S}(t)=(S_1(t), S_2(t), \ldots, S_n(t))^{\top}$ and define $\mathbf{H}(t), \mathbf{C}(t), \mathbf{R}(t), \mathbf{C}^{new}(t) \in\mathbb{R}^n$ similarly. We only observe $\mathbf{N}$ and $\mathbf{C}^{new}(t)$ at time $t$.

We consider the scenario where the testing capacity is limited: At each time $t$, we can only actively test $M$ individuals for all counties. An {\it active screening strategy} decides how many individuals to test for each county. It is represented by an allocation vector $\mathbf{a}(t)=(a_1(t), a_2(t),\ldots,a_n(t))^{\top}$, where $N_ia_i(t)$ individuals of county $i$ will be tested at time $t$, $1\leq i\leq n$. We call $a_i(t)$ the {\it testing rate} of county $i$ at time $t$. 
This testing rate will affect the number of confirmed cases at $(t+1)$. This is captured by an ordinary difference equation model to be introduced in Section~\ref{subsec:model}. 

Suppose testing is not available during $t=1,2,\ldots,t_0-1$ (a period where the disease spreads but there is no intervention). After that period, a capacity of $M$ tests is available at every $t=t_0, t_0+1,\ldots, T$. At time $(t_0-1)$, the administrator has to decide the allocation vectors $\mathbf{a}(t_0), \mathbf{a}(t_0+1),\ldots,\mathbf{a}(T)\in\mathbb{R}^n$ in order to minimize the total number of hidden cases in all counties during $t_0$ to $T$:
\begin{equation} \label{Problem}
\min_{\mathbf{a}(t_0: T)} \Bigl\{ \sum_{t=t_0}^T\sum_{i=1}^n H_i(t)\Bigr\}, \ \mbox{subject to:}\  \sum_{i=1}^n N_i(t)a_i(t)\leq M, \  \mbox{for every $t_0\leq t\leq T$}. 
\end{equation}
The data available to the administrator at $(t_0-1)$ are $\mathbf{N}$ and $\{\mathbf{C}^{new}(t)\}_{1\leq t\leq t_0-1}$. The administrator will first use these historical data to learn the patterns of disease transmission (i.e., to estimate model parameters) and then use the estimated model to find the optimal allocation vectors.

Below, in Section~\ref{subsec:model}, we introduce the disease model, where a key ingredient is using the commute network of counties to help model disease transmission. In Section~\ref{subsec:optimization}, we describe how to find the optimal allocation vectors when model parameters are known. In Section~\ref{subsec:estimation}, we describe how to estimate model parameters.

\subsection{A network SIR model with the screening intervention} \label{subsec:model}

For any $i\neq j$, let $L_{ij}>0$ be the traffic distance between county $i$ and county $j$. We define a (weighted) commute network with $n$ nodes, where each node represents a county, and the weighted edges between two counties are 
\begin{equation} \label{Mod-network}
\mbox{$w$}_{ij} = \begin{cases}
\lambda /L^2_{ij}, & \mbox{if }i\neq j,\cr
1-\sum_{k\neq i}w_{ik} & \mbox{if }i=j.
\end{cases}
\end{equation}
We use this network to describe the cross-county interactions. The larger $w_{ij}$, the more interactions between individuals of counties, and the easier transmission of disease. 
The parameter $\lambda>0$ controls the level of travel restrictions; e.g., a `lockdown' policy of all counties means $\lambda=0$.

We recall that $N_i$ is the population of county $i$. 
Let $s_i(t)=S_i(t)/N_i$, $h_i(t)=H_i(t)/N_i$, $c_i(t)=C_i(t)/N_i$ and $r_i(t)=R_i(t)/N_i$, for $1\leq i\leq n$ and $1\leq t\leq T$. It suffices to model these ratios $(s_i(t), h_i(t), c_i(t), r_i(t))$. We impose an ordinary difference equation model. 
The model has three parameters, the infection rate $\alpha>0$, the diagnosis rate $\beta>0$ and the recovery rate $\gamma>0$. Fixing a county $i$, from time $t$ to time $(t+1)$, we assume that a $\gamma$ fraction of the infected cases are recovered, which gives $r_i(t+1)-r_i(t)=\gamma[h_i(t)+c_i(t)]$. We also assume that the fraction of hidden cases getting confirmed is equal to $
\beta + a_i(t)$, where $a_i(t)$ is the testing rate allocated to county $i$. It follows that $c_i(t+1)-c_i(t)=[\beta + a_i(t)] h_i(t) - \gamma c_i(t)$. Furthermore, let $p_i(t)$ denote the probability of a susceptible individual getting infected. We assume
\begin{equation} \label{Mod-InfectRate}
p_i(t) = p_i(\mathbf{h}(t), \mathbf{W}) = 1 - \prod_{j=1}^n \bigl[1 - \alpha w_{ij}h_j(t)\bigr],
\end{equation}
where $\mathbf{h}(t)=(h_1(t), h_2(t),\ldots,h_n(t))^{\top}$ and $\mathbf{W}=(w_{ij})$ is the adjacency matrix of the commute network. $p_i(t)$ only depends on the fractions of hidden cases, but not the fractions of confirmed cases. This is because we make an (idealized) assumption that each confirmed case is properly quarantined and will not infect other susceptible individuals. The commute network plays a role in the disease spread across counties. If county $i$ is isolated (i.e., $w_{ij}=0$ for all $j\neq i$), then $p_i(t)=\alpha h_i(t)$. When county $i$ is not isolated, $p_i(t)$ is affected by the $h_j(t)$ of all counties connected to node $i$ in the commute network. Using this notation of $p_i(t)$, we further have $s_i(t+1) - s_i(t) = - p_i(t) s_i(t)$ and $h_i(t+1) - h_i(t) = p_i(t) s_i(t) - [\beta + a_i(t)] h_i(t) - \gamma h_i(t)$. 
This model is summarized in the following set of difference equations:
\begin{equation} \label{Mod-ODE}
    \left\{ \begin{split}
\Delta s_i(t) &= -p_i(t) s_i(t) \\
\Delta h_i(t) &=p_i(t) s_i(t) - [\beta + a_i(t)+\gamma ] h_i(t) \\
\Delta c_i(t) &= [\beta + a_i(t)] h_i(t) - \gamma c_i(t)\\
\Delta r_i(t) &= \gamma\, [h_i(t)+c_i(t)]
\end{split} \right., \  \mbox{for every }1\leq i\leq n. 
\end{equation}

In the case that there is no intervention (i.e., $\mathbf{a}_t$ is a zero vector) and that all counties are isolated (i.e., $\mathbf{W}$ is a diagonal matrix), this model reduces to a standard 4-compartment SIR model for every county. 
When there is no intervention, the way we incorporate the effect of the commute network is similar to that in a network-based compartmental model \citep{wang2003epidemic}. However, a standard network-SIR model is defined on the contact tracing network, where each node is an individual, and the state of a node at $t$ is a categorical variable describing which compartment this individual belongs to. In comparison, our model is defined on the commute network, where each node is a county, and the state of a node is characterized by a vector $(s_i(t), h_i(t), c_i(t), r_i(t))$.  The way we incorporate the effect of intervention is similar to the model used in active screening \citep{ou2020and}. The model in \cite{ou2020and} is for the SIS disease and on the contact tracing network of individuals, while our model is for the SIR disease and on the commute network of counties.

\subsection{Optimization of allocation vectors} \label{subsec:optimization}

When the commute network $\mathbf{W}$ is given and when the model parameters $(\alpha,\beta,\gamma)$ are known, we aim to solve \eqref{Problem} to find the optimal allocation vectors $\mathbf{a}(t_0), \mathbf{a}(t_0+1),\ldots,\mathbf{a}(T)$. Recall that $\mathbf{h}(t)=(h_1(t), h_2(t),\ldots,h_n(t))^{\top}$ and $\mathbf{N}=(N_1,N_2,\ldots,N_n)^{\top}$. We re-write \eqref{Problem} as
\begin{equation} \label{Obj-1}
\min_{\mathbf{a}(t_0:T)} \mbox{$F$}(\mathbf{a}(t_0:T)):={\bf N}^{\top}\bigl[\sum_{t=t_0}^T \mathbf{h}(t)\bigr], \ \mbox{subject to:}\quad \mathbf{N}^{\top}\mathbf{a}(t) \leq M, \  \mbox{for every $t_0\leq t\leq T$}. 
\end{equation}
Since $\Delta h_i(t)$ depends on the allocation vectors in a complicated way as specified by models \eqref{Mod-InfectRate}-\eqref{Mod-ODE}, it is not easy to solve this optimization directly.
We borrow the idea in \cite{ou2020and} to minimize a different objective, $F^*(\mathbf{a}(t_0:T))$, which is an (approximate) upper bound of $F(\mathbf{a}(t_0:T))$.

By model \eqref{Mod-ODE}, for every $1\leq i\leq n$ and $t_0-1\leq t\leq T$, 
\begin{equation} \label{simplify-1}
h_i(t+1)=h_i(t) [1- \beta -\gamma - a_i(t)]  +p_i(t) s_i(t).
\end{equation}
To simplify \eqref{simplify-1}, first, note that we are interested in the early period of disease progression when the number of individuals in compartments $H$, $C$ and $R$ are negligible compared with the number of individuals in $S$. It yields $s_i(t)=S_i(t)/N_i\approx 1$. This motivates us to set $s_i(t)\equiv 1$ for simplicity. 
Next, we consider the expression of $p_i(t)$ in \eqref{Mod-InfectRate}. 
We apply the universal inequality $1-\alpha z - (1-\alpha)^z\geq 0$, for any $\alpha\in (0,1)$ and $z>0$. 
It implies $p_i(t) = 1 - \prod_{j=1}^n \bigl[1 - \alpha w_{ij}h_j(t)\bigr]\leq 1-\prod_{j=1}^n (1-\alpha)^{w_{ij}h_j(t)} = 1-(1-\alpha)^{\sum_{j=1}^n w_{ij}h_j(t)}$.  When $\alpha$ is sufficiently small, we apply the Taylor expansion to $g(\alpha) = (1-\alpha)^{\sum_{j=1}^n w_{ij}h_j(t)}$ at the origin. It gives $g(\alpha)\approx 1-\alpha\sum_{j=1}^n w_{ij}h_j(t)$. We combine the above to get
\begin{equation} \label{Proxy-p_i(t)}
p_i(t) \leq 1-(1-\alpha)^{\sum_{j=1}^n w_{ij}h_j(t)}\approx \alpha\sum_{j=1}^n w_{ij}h_j(t). 
\end{equation}
We plug \eqref{Proxy-p_i(t)} and $s_i(t)=1$ into \eqref{simplify-1}. It yields a proxy inequality, $h_i(t+1) \lesssim  h_i(t) [1- \beta -\gamma - a_i(t)]  +\sum_{j}  \alpha w_{ij}h_j(t)$, which can be re-written as
\begin{eqnarray} \label{Proxy-h_i(t)}
h_i(t+1) 
&\leq & \mathbf{e}_i^{\top}\mathbf{U}(t) \mathbf{h}(t), \ \mbox{where}\  \mathbf{U}(t) := (1- \beta -\gamma)\mathbf{I}_n - \mathrm{diag}(\mathbf{a}(t)) +\alpha \mathbf{W}. 
\end{eqnarray}
By stacking \eqref{Proxy-h_i(t)} for $1\leq i\leq n$,  we obtain its matrix form as $\mathbf{h}(t+1)\leq \mathbf{U}(t) \mathbf{h}(t)$, where `$\leq $' is in the entry-wise fashion. In the parameter range of interest, $\beta,\gamma$ and $a_i(t)$ are all much smaller than $1$. Therefore, the matrix $\mathbf{U}(t)$ is always a nonnegative matrix. We can iterate this inequality to get $\mathbf{h}(t)\leq\bigl[ \prod_{m=t_0-1}^{t-1}\mathbf{U}(t)\bigr]\mathbf{h}_0$, where $\mathbf{h}_0=\mathbf{h}(t_0-1)$ and  `$\leq $' in the entry-wise fashion. It yields a relaxation of \eqref{Obj-1} as
\begin{equation} \label{Obj-2}
\min_{\mathbf{a}(t_0:T)} \mbox{$F$}^*(\mathbf{a}(t_0:T)):= \mathbf{N}^{\top}\biggl(\sum_{t=t_0}^T \bigl[ \prod_{m=t_0-1}^{t-1}\mathbf{U}(m)\bigr]\biggr)\mathbf{h}_0, \ \mbox{subject to:}\   \mathbf{N}^{\top}\mathbf{a}(t) \leq M, \  \mbox{for every $t_0\leq t\leq T$}. 
\end{equation}

We propose to compute the allocation vectors by solving \eqref{Obj-2}. The gradient of $F^*(\mathbf{a}(t_0:T))$ has a nice form. Let $\nabla_t F^*$ be the gradient of $F^*$ with respect to $\mathbf{a}(t)$. By direct calculations, 
\begin{equation} \label{Gradient}
\nabla_t \mbox{$F$}^* =  - \sum_{m= t+1}^T \bigl[ \mathbf{U}(t+1)\mathbf{U}(t+2)\ldots \mathbf{U}(m) \mathbf{N} \bigr]\circ 
   \bigl[ \mathbf{U}(t_0)\mathbf{U}(t_0+1)\ldots \mathbf{U}(t-1) \mathbf{h}_0 \bigr]. 
\end{equation}
We apply the Frank-Wolfe algorithm. Let $\mathbf{a}^{0}(t_0:T)$ be the initial solution. At iteration $(k+1)$, given $\mathbf{a}^{k}(t_0:T)$, we compute $\mathbf{a}^{k+1}(t_0:T)$ as follows:
\begin{enumerate}
\item Compute $\nabla_t F^*$ at $\mathbf{a}^k(t_0:T)$. Solve $\mathbf{b}(t_0),\ldots,\mathbf{b}(T)$ by minimizing $\sum_{t=t_0}^T [\mathbf{b}(t)]^{\top}\bigl[\nabla_t F^*(\mathbf{a}^k(t_0:T))\bigr]$, subject to the constraint $\mathbf{N}^{\top}\mathbf{b}(t)\leq M$ for every $t_0\leq t\leq T$. This is a standard linear programming, which we solve using Dantzig's simplex method. 
\item Update $\mathbf{a}^{k+1}(t)=\mathbf{a}^k(t) + (k+1)^{-1}[\mathbf{b}(t) - \mathbf{a}^k(t)]$, for every $t_0\leq t\leq T$. 
\end{enumerate}

The input of the algorithm are the adjacency matrix $\mathbf{W}$ of the commute network, model parameters $(\alpha,\beta,\gamma)$, and the vector $\mathbf{h}_0$ that contains the fractions of hidden cases at $(t_0-1)$ for all counties. In the next subsection, we describe how to obtain these input from historical data.

\subsection{Estimation of model parameters} \label{subsec:estimation}
At time $(t_0-1)$, the administrator observes the historical data of newly confirmed cases $\{\mathbf{C}^{new}(t)\}_{1\leq t\leq t_0-1}$. Additionally, the county-wise populations $\{N_i\}_{1\leq i\leq n}$ and the between-county traffic distances $\{L_{ij}\}_{1\leq i<j\leq n}$ are known. To run the algorithm in Section~\ref{subsec:optimization}, we need to know $(\alpha,\beta,\gamma,\lambda)$ and $\mathbf{h}_0=\mathbf{h}(t_0-1)$, where $\lambda$ is the parameter in \eqref{Mod-network} for defining the commute network. 
We first estimate $(\beta,\gamma)$ from knowledge of the disease; next, we estimate $(\alpha,\lambda)$ by fitting a network SIR model without intervention; last, we estimate $\mathbf{h}_0$ directly from other estimated parameters. 

For a SIR disease, the diagnosis rate $\beta$ and the recovery rate $\gamma$ are related to the nature of the virus, thus invariant with locations or time. Since we only observe the number of newly confirmed cases, it is infeasible to estimate $(\beta,\gamma)$ from data. We follow the convention to estimate them from clinical records. Let $d_r$ be the average time for an infected person to get recovered. It is common to estimate $\gamma$ by $1/d_r$. To get an estimate of the diagnosis rate $\beta$, we note that there are two kinds of infected cases: One is the asymptomatic case, who never shows symptoms. We assume that these individuals will never be diagnosed without screening interventions (but they are infectious during the infected period). The other is the symptomatic case, who have symptoms after an `incubation' period (in the incubation period, they have no symptoms but are still infectious). Let $\delta_a$ be the percent of asymptomatic cases, and let $d_h$ be the average time of the incubation period for symptomatic cases. We estimate $(\beta,\gamma)$ by
\begin{equation} \label{Estimate-(beta+gamma)}
\hat{\beta} = (1-\delta_a)/d_h \  \mbox{and}\  \hat{\gamma}=1/d_r. 
\end{equation}
For COVID-19, there have been many medical literature that provide information of $(\delta_a, d_r, d_h)$. We set $\delta_a=0.17$ following \cite{Byambasuren2020}, $d_h = 5.2$ days following \cite{Li2020}, and $d_r = 15$ days following \citet{Beigel2020}. It gives $\hat{\beta}=0.16$ and $\hat{\gamma}=0.067$ for COVID-19.

We then estimate $(\alpha, \lambda)$ given $(\hat{\beta},\hat{\gamma})$. During the time period of $1,2,\ldots,t_0-1$, the transmission of disease follows a natural process without screening intervention. We use a special case of models \eqref{Mod-InfectRate}-\eqref{Mod-ODE} with $a_i(t)\equiv 0$. The difference equation model becomes
\begin{equation} \label{Mod-ODE-noScr}
    \left\{ \begin{split}
\Delta s_i(t) &= -p_i(t) s_i(t) \\
\Delta h_i(t) &=p_i(t) s_i(t) - [\beta+\gamma ] h_i(t) \\
\Delta c_i(t) &= \beta h_i(t) - \gamma c_i(t)\\
\Delta r_i(t) &= \gamma\, [h_i(t)+c_i(t)]
\end{split} \right., \  \mbox{where}\  \mbox{$p_i(t) = 1 - \prod_{j=1}^n \bigl[1 - \alpha w_{ij}h_j(t)\bigr]$}. 
\end{equation}
The available data are numbers of newly confirmed case $\{C^{new}_{i}(t)\}_{1\leq i\leq n,1\leq t\leq t_0-1}$. By model \eqref{Mod-ODE-noScr}, $C_i^{new}(t+1)=\beta N_ih_i(t)$. It follows that
\begin{equation} \label{Estimate-(alpha+lambda)-1}
\mbox{$\hat{h}_i(t) = (N_i\hat{\beta})^{-1}C_i^{new}(t+1)$}, \  0\leq t\leq t_0-2. 
\end{equation}
We introduce two approaches for the estimation of $(\alpha,\lambda)$. Write $p_i(t)=p_i(t,\alpha,\lambda)$ to indicate its dependence on the unknown parameters. 
In the first approach, we note that when $s_i(t)\approx 1$, by model \eqref{Mod-ODE-noScr}, $ \Delta [\ln(h_i(t))] \approx [\Delta h_i(t)]/h_i(t) \approx p_i(t, \alpha, \lambda)/h_i(t) - (\beta+\gamma)$. 
Therefore, at each $t$, we make a 1-step ahead forecast of $\ln(h_i(t))$ by $\ln(\hat{h}_i(0))+\sum_{\tau=0}^{t-1}\bigl[\hat{p}_i(\tau, \alpha,\lambda)/\hat{h}_i(\tau)-(\hat{\beta}+\hat{\gamma})\bigr]$, where $\hat{p}_i(\tau,\alpha, \lambda)=1 - \prod_{j=1}^n [1 - \alpha w_{ij}(\lambda) \hat{h}_j(\tau)]$ and $w_{ij}(\lambda)$ is as in \eqref{Mod-network}. We estimate $(\alpha,\lambda)$ by minimizing the sum of squared forecast errors:
\begin{equation} \label{Estimate-(alpha+lambda)-2}
(\hat{\alpha}, \hat{\lambda}) = \mathrm{argmin}_{(\alpha,\lambda) } \Bigl\{ \mbox{$\sum_{t=0}^{t_0-2}\sum_{i=1}^n \bigl[\ln(\hat{h}_i(t)) - \ln(\hat{h}_i(0)) + (\hat{\beta}+\hat{\gamma})t - \sum_{\tau=0}^{t-1}\hat{p}_i(\tau, \alpha, \lambda)/\hat{h}_i(\tau)\bigr]^2$}\Bigr\}. 
\end{equation}
In the special case of $\mathbf{W}=\mathbf{I}_n$, it holds that $\hat{p}_i(\tau, \alpha, \lambda)=\alpha \hat{h}_i(\tau)$, and the above reduces to a least-squares of $\ln (\hat{h}_i(t))$ versus $t$, which agrees with the method in \cite{Becker1976} for a standard SIR model. In the second approach, given any $(\alpha,\lambda)$, we can use $\hat{\mathbf{h}}(0)$ to forecast $\mathbf{h}(1:t_0-2)$ based on model \eqref{Mod-ODE-noScr}: We apply the recursive formula of $\hat{h}^{forecast}_i(t+1)= \hat{p}_i(t,\alpha,\lambda)+(1-\hat{\beta}-\hat{\gamma})\hat{h}_i^{forecast}(t)$, where $\hat{p}_i(t,\alpha,\lambda)$ is the same as above. This produces $\hat{\mathbf{h}}^{forecast}(1),\hat{\mathbf{h}}^{forecast}(2),\ldots,\hat{\mathbf{h}}^{forecast}(t_0-1)$ recursively. We minimize the total forecast errors: 
\begin{equation} \label{Estimate-(alpha+lambda)-3}
(\hat{\alpha}, \hat{\lambda}) = \mathrm{argmin}_{(\alpha,\lambda) } \Bigl\{ \mbox{$\sum_{i=1}^n \bigl| \sum_{t=1}^{t_0-2} N_i \hat{h}_i(t) - \sum_{t=1}^{t_0-2} N_i \hat{h}^{forecaset}_i(t) \bigr|$} \Bigr\}. 
\end{equation}
The two approaches both have reasonably good performances in simulations.

Last, we estimate $\mathbf{h}_0=\mathbf{h}(t_0-1)$. Since \eqref{Estimate-(alpha+lambda)-1} gives $\hat{\mathbf{h}}(0),\hat{\mathbf{h}}(1),\ldots,\hat{\mathbf{h}}(t_0-2)$, we estimate $\mathbf{h}(t_0-1)$ by the 1-step ahead forecast:
\begin{equation}\label{Estimate-h0}
\mbox{$\hat{\mathbf{h}}(t_0-1) = (1-\hat{\beta}-\hat{\gamma})\hat{\mathbf{h}}(t_0-2)+\hat{\mathbf{p}}(t_0-2)$}, \ \mbox{where}\  \mbox{$\hat{p}_i(t_0-2) = 1- \prod_{j = 1}^n \bigl[ 1 - \hat{\alpha}w_{ij}(\hat{\lambda})\hat{h}_j(t_0-2)\bigr]$}. 
\end{equation}

\section{Semi-synthetic experiments on real networks} \label{sec:RealData}

We conduct semi-synthetic experiments on real commute networks to test the performance of our method. We consider the state of Massachusetts in the United States and the province of Hubei in People's Republic of China. 
The traffic distance data came from Google Map and Bing Map (for Massachusetts) and Amap (for Hubei). We searched for the average driving distances between counties on these map services and used them to build the commute network as in  \eqref{Mod-network}. We also used the data of daily confirmed COVID-19 cases from the COVID-19 dashboard of Johns Hopkins University (for Massachusetts) and the National Health Commission of People's Republic of China (for Hubei). The code of all numerical experiments can be found at \url{https://github.com/ZhengTracyKe/COVID-Screening}.

\begin{figure}[tb!]
\centering
\includegraphics[width=0.4\linewidth]{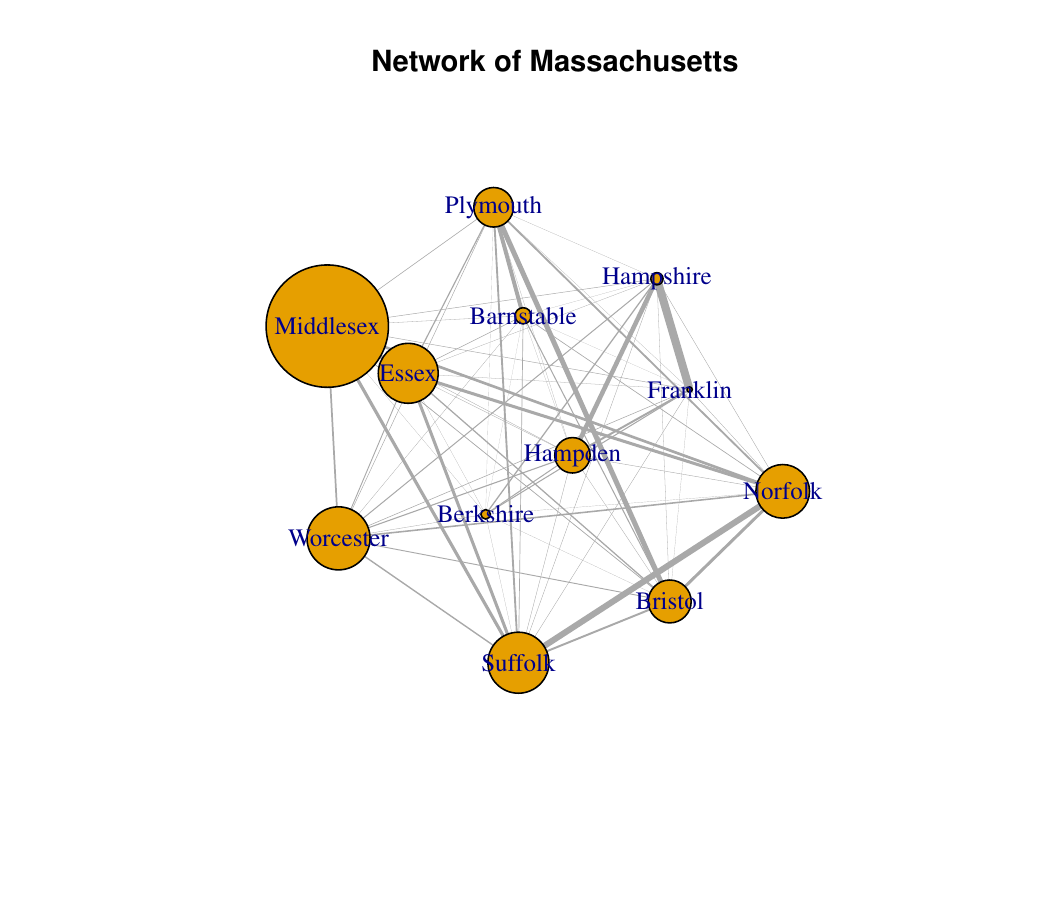}
\includegraphics[width=0.4\linewidth]{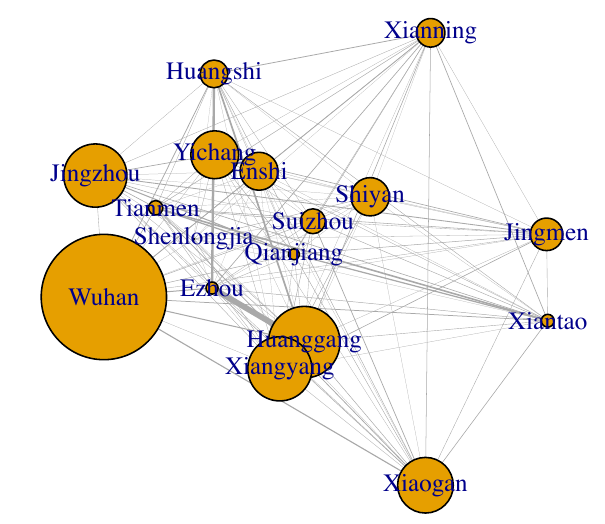}
\caption{The commute networks of Massachusetts (left) and Hubei (right). The size of a node is proportional to the population. The width of an edge is proportional to the reciprocal of squared distance between two nodes. For the network of Massachusetts, Dukes and Nantucket counties are removed, because their number of confirmed cases in real data are negligible compared with other counties. The maps of Massachusetts and Hubei are in the supplemental material.} \label{map} 
\end{figure}

Hubei is the most severely infected province in China during February 2020, with a total of 49,497 confirmed cases on Feb. 29th. Wuhan, the capital city of Hubei, reported the first COVID-19 case in December 2019. Hubei has a population of around 58.5 million. It contains one big capital city, Wuhan, several middle-size cities like Huanggang, Xiangyang and Jingzhou, and some small provincial administered county-level divisions like Tianmen, and Shenlongjia (see Figure~\ref{map}). 
Massachusetts is a state on the east coast of U.S. with a population of around 6.9 million, and had a total of 757,849 confirmed cases by Sep. 30, 2021. 
Compared with the commute network of Hubei, the commute network of Massachusetts has less severe degree heterogeneity, and the population is also more evenly distributed among counties (see Figure~\ref{map}).

\subsection{Comparison of allocation strategies} \label{subsec:simu-compare}

We use the traffic distances $\{L_{ij}\}_{1\leq i<j\leq n}$ and population sizes $\{N_i\}_{1\leq i\leq n}$ from real data, and simulate $\{(S_i(t), H_i(t), C_i(t), R_i(t))\}_{1\leq i\leq n}$ using the model in Section~\ref{subsec:model}, with various choices of parameters $(\alpha,\beta,\gamma, \lambda, M)$. 
We fix $t_0=1$ and $T=30$.  The initial numbers of individuals in four compartments are set as follows: we pick a day at the early stage of the pandemic as $t_0$ and obtain $N_i$, $C_i(t_0-1)$, $C_i^{new}(t_0-1)$ and $C_i^{new}(t_0)$ from real data; we then set $H_i(t_0-1)=\beta^{-1}C_i^{new}(t_0)$, $S_i(t_0-1)=N_i$ and $R_i(t_0-1)=0$. 
In this subsection, we assume that the true parameters $(\alpha,\beta,\gamma,\lambda)$ are given and focus on evaluating the performance of the allocation strategy. Parameter estimation will be investigated in next subsection. 

We compare our screening strategy with two other strategies: (a) Allocation by population: it sets a flat testing rate for all counties, so that the testing budget allocated to a county is proportional to its population. This strategy guarantees fairness, but it is less satisfactory for controlling the spread of disease. (b) Allocation by infection rate: it sets the testing rate $a_i(t)$ to be proportional to $C_i^{new}(t-1)/N_i$. This strategy takes into account the difference of infection rates among counties, but it does not utilize the commute network. Additionally, we include the `no screening' strategy as a reference. For each strategy, we measure its performance by tracing the number of cumulative confirmed cases, $C^{cum}(t)=\sum_{\tau=t_0}^t\sum_{i=1}^n C_i^{new}(\tau)$. 
Since only newly confirmed cases are reported in reality, this performance metric is natural, and it makes the comparison with real data easy. We note that this metric is not equivalent to counting the total number of hidden cases, and so it does not automatically favor our method.

\begin{figure}[tb!]
\centering
\includegraphics[width=0.8\linewidth]{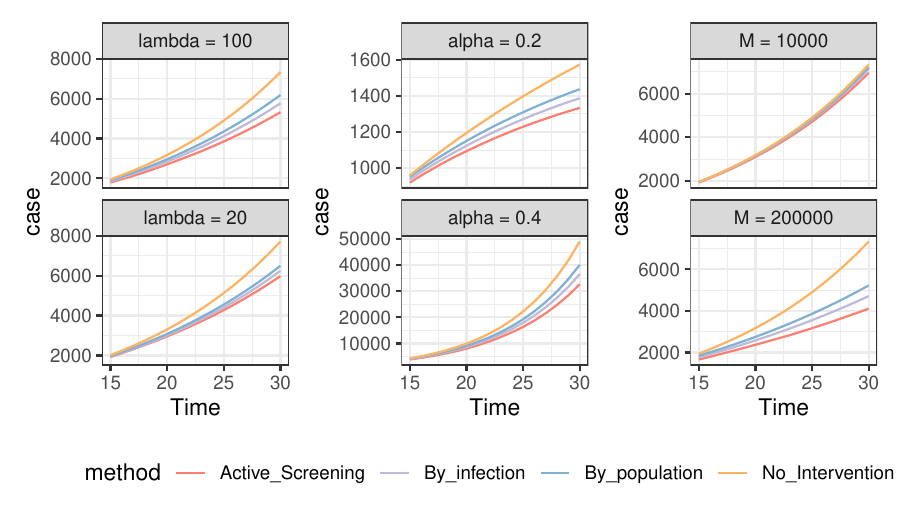}
\includegraphics[width=0.8\linewidth]{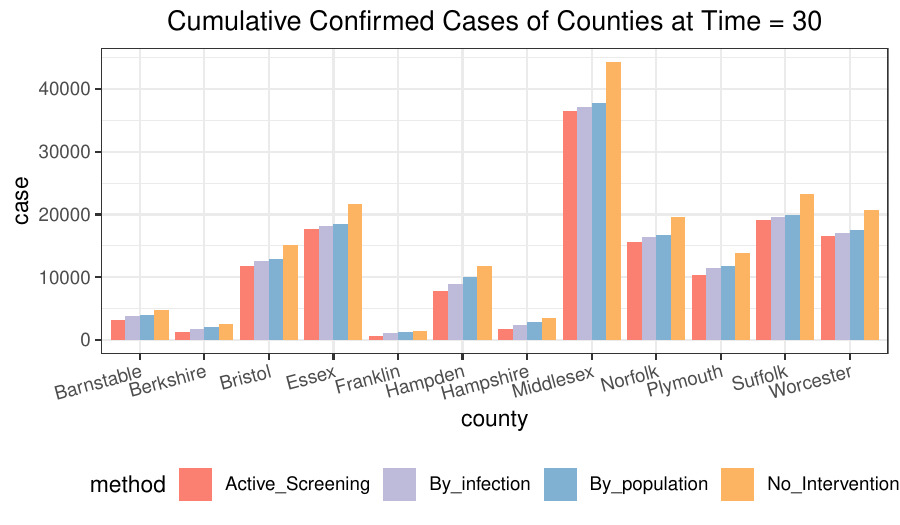}
\caption{Experiments for Massachusetts. Left: total number of cumulative confirmed cases from day 15 to day 30. Right:  number of cumulative confirmed cases on day 30 for each county. Unless noted in plots, the default parameters are: $\lambda$ = 100, $\beta=0.16$, $\gamma = 1/15$,  $\alpha$ = 0.3, M = 100K.} \label{simulation_massachusetts}
\end{figure}

{\bf The Massachusetts network}.  Massachusetts has 14 counties. Since Dukes and Nantucket reported very few confirmed cases in the early stage of the pandemic, for simplicity, we remove these two counties and consider the commute network of 12 counties. In the default parameter setting, we let $(\alpha,\beta,\gamma,\lambda, M)= (0.3, 0.16, 1/15, 100 \text{mi}, 100K)$. The default diagnosis rate $\beta$ and the default recovery rate $\gamma$ are from the estimates based on clinical data of COVID-19 (see Section~\ref{subsec:estimation}). We vary the other parameters $(\alpha,\lambda, M)$. The resulting $C^{cum}(t)$ of different strategies are shown in Figure~\ref{simulation_massachusetts} (left). In all settings considered here, our allocation strategy outperforms the other strategies. The parameter $\lambda$ controls the weights in the commute network. We consider $\lambda\in \{20\text{mi},100\text{mi}\}$ (other parameters take the default values; same below). The larger $\lambda$, the more impact of the network structure on the pandemic progression, hence, the more advantage of our network-aware strategy. We also consider $\alpha\in \{0.2, 0.3, 0.4\}$ (the results for $\alpha=0.3$ are in the top left plot, and the results for $\alpha\in \{0.2, 0.4\}$ are in the middle plots). The curve $C^{cum}(t)$ has different behaviors as $\alpha$ varies: it has a super-linear growth for $\alpha\in\{0.3, 0.4\}$ and a sub-linear growth for $\alpha=0.2$. In both cases, our strategy performs the best, and the no-screening strategy performs the worst; the strategy of allocation by infection rate is better than the strategy of allocation by population. We also consider $M\in\{10K, 100K, 200K\}$ (the results for M=100K are in the top left plot, and the results for $M\in \{10K, 200K\}$ are in the right plots). Massachusetts has a population of 6.9 million. A daily testing budget of 10K has a small effect on the pandemic progression, so the performances of different strategies are pretty close. When the daily budget is increased to 200K, the effect of the screening strategy becomes significant: the number of cumulative confirmed cases on day 30 by our strategy is only half of the number associated with no screening. 

\begin{figure}[tb!]
\centering
\includegraphics[width=0.9\linewidth]{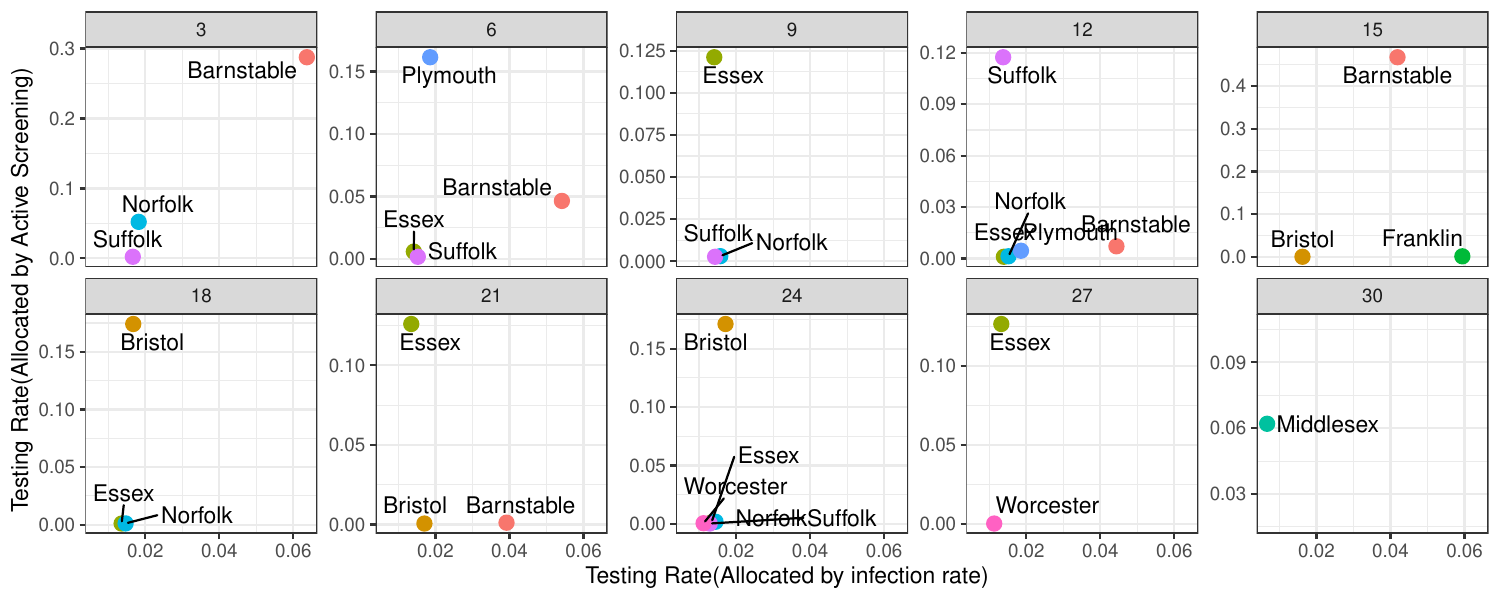}
\caption{Comparison of the allocated testing rate by our method versus the infection rate. We only plot counties with a non-zero testing rate in our method. The parameters are $\lambda$ = 100, $\beta=0.16$, $\gamma = 1/15$,  $\alpha$ = 0.3, M = 100K.} \label{allocation_massachusetts}
\end{figure}

In Figure~\ref{simulation_massachusetts} (right), we plot the cumulative number of confirmed cases on day 30 for each county, when parameters take the aforementioned default values. Interestingly, although our method is not a `fair' screening strategy (non-flat testing rate), the benefit is `universal' --- compared with other strategies, there is a decrease of confirmed cases for all counties. We note that for some counties, they get lower allocation rates than they would have got from allocation by population or infection rate, but their confirmed cases are still reduced. It suggests that, to reduce the confirmed cases at one county, it is sometimes more effective to increase the testing rate of nearby highly infected counties than increasing the testing rate of this county. This explains why it is useful to take into account the network structure. 
In Figure \ref{allocation_massachusetts}, we take a close look at the allocation vectors. 
For each $t\in \{3,6,9,\ldots,27, 30\}$, we plot the allocated testing rate $a_i(t)$ versus the infection rate of the same day. Only counties with a nonzero $a_i(t)$ are shown in the plots. 
Our screening strategy is `sparse': Each day, it puts all testing budget to a small number of counties. This is different from allocation by population or infection rate, where almost every county gets some testing resource each day. For all 30 days, the most frequently selected counties are Essex (17 days), Suffolk (17 days) and Norfolk (15 days). Norfolk and Suffolk have the highest degrees in the commute network, and Essex is close to Suffolk and Middlesex (the largest-population county). Our screening strategy tends to select counties with a high infection rate (e.g., Barnstable, Berkshire) in early periods and counties with a large population (e.g., Middlesex, Worcester) in the late periods.

{\bf The Hubei network}. Hubei has 17 cities, including 12 prefecture-level cities, 1 autonomous prefecture and 4 provincial administrated county-level cities. 
Compared with the Massachusetts network, the Hubei network has several special characteristics. First, the capital city, Huhan, plays a dominating role. Wuhan has far more confirmed cases than other cities at the beginning of the pandemic. Wuhan also has the largest population. Second, the population of Hubei is nearly 8 times of the population of Massachusetts. Given the same testing budget (e.g., 100K), the effect of active screening is less significant for Hubei, and the allocation vector is also sparser.   
Third, the Hubei network has more severe degree heterogeneity. We use $\ell_i\equiv \sum_{j:j\neq i}w_{ij}\in [0,1]$ to measure the degree of a node. In the Hubei network, when $\lambda=50$, the three largest $\ell_i$'s are 0.848 (Ezhou), 0.797 (Huanggang) and 0.217 (Huangshi). Ezhou and Huanggang are hub nodes, having much larger degrees than the other nodes. 
In comparison, the three largest $\ell_i$'s for Massachusetts is 0.652 (Norfolk), 0.636 (Suffolk) and 0.556 (Hampshire), there is no clear hub node. 
Due to these special characteristics, the simulation results for Hubei are different from those for Massachusetts.

\begin{figure}[tb!]
\centering
\includegraphics[width=0.8\linewidth]{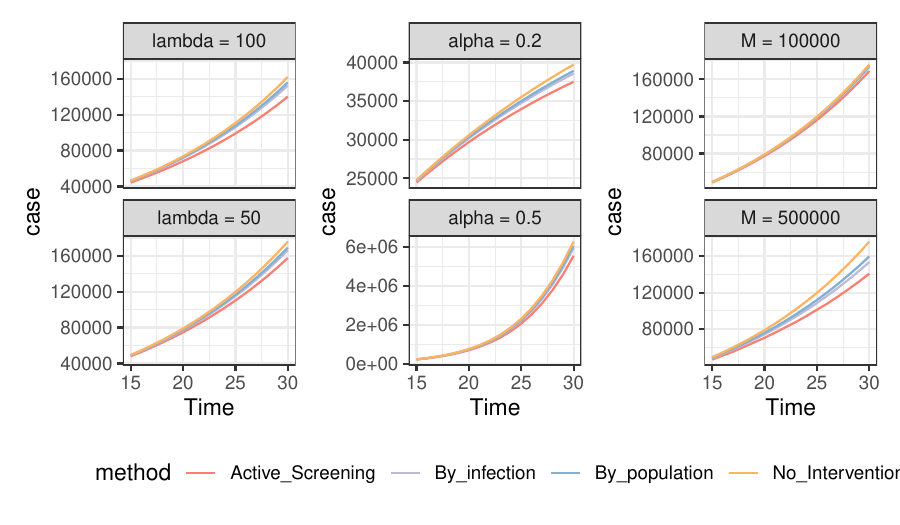}
\includegraphics[width=0.8\linewidth]{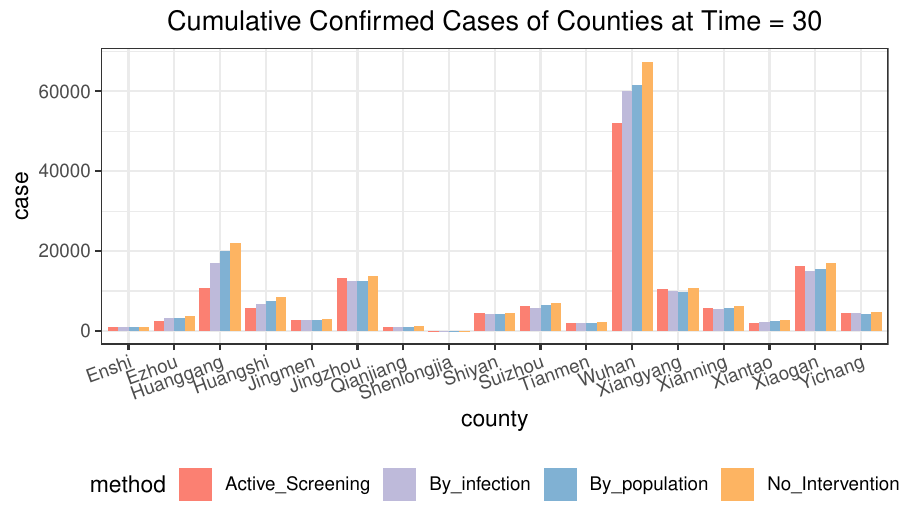}
\caption{Experiments for Hubei. Left: total number of cumulative confirmed cases from day 15 to day 30. Right:  number of cumulative confirmed cases on day 30 for each county. Unless noted in plots, the default parameters are: $\lambda$ = 50, $\beta=0.16$, $\gamma = 1/15$,  $\alpha$ = 0.3, M = 200K.} \label{simulation_hubei}
\end{figure}

\begin{figure}[tb!]
\centering
\includegraphics[width=0.9\linewidth]{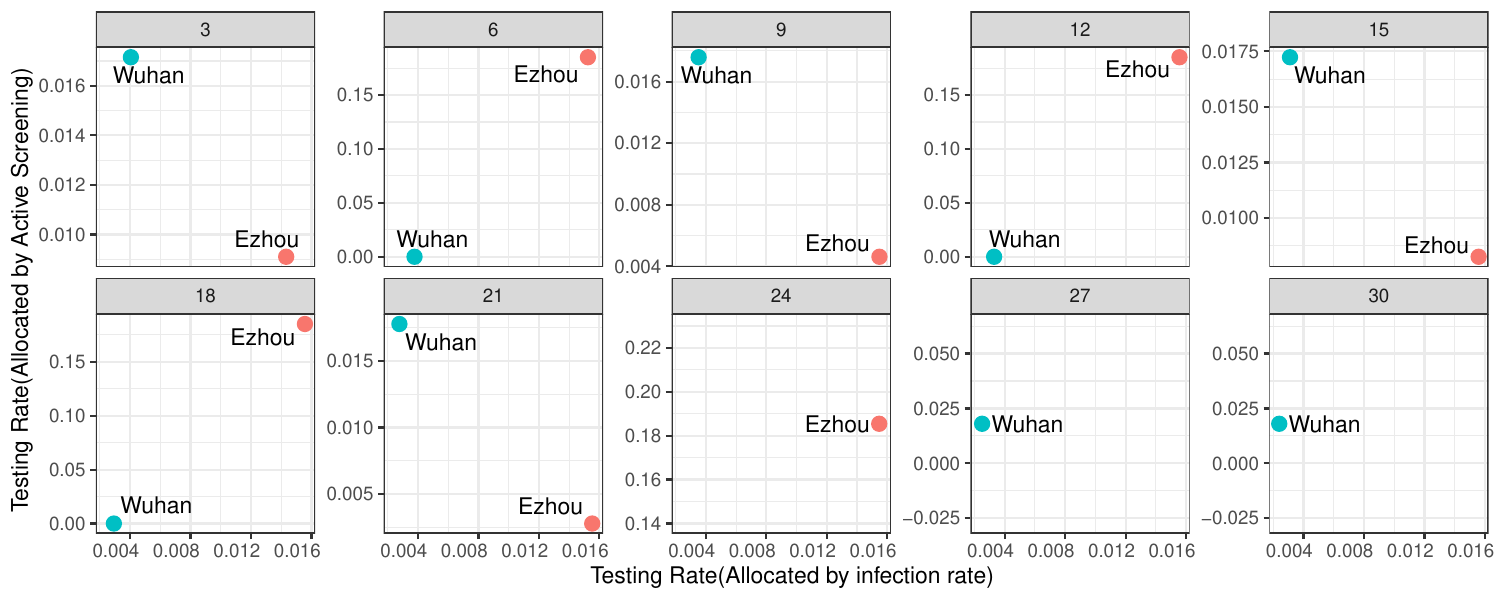}
\caption{Comparison of the allocated testing rate by our method versus the infection rate. We only plot counties with a non-zero testing rate in our method. The parameters are $\lambda$ = 50, $\beta=0.16$, $\gamma = 1/15$,  $\alpha$ = 0.3, M = 200K.} \label{allocation_hubei_1}
\end{figure}

We set the default parameters as $(\alpha,\beta,\gamma,\lambda, M)= (0.3, 0.16, 1/15, 50 \text{mi}, 200K)$. We also consider $\lambda\in \{50 \text{mi}, 100 \text{mi}\}$, $\alpha\in \{0.2, 0.3, 0.5\}$ and $M\in \{100K, 200K, 500K\}$ (every time we vary one parameter and fix the other parameters as the default values). The results are in Figure~\ref{simulation_hubei} (left). 
It suggests that our screening strategy outperforms all the competitors. 
Since Hubei has a much larger population and more confirmed cases than Massachusetts, the improvement relative to no screening is smaller; however, the absolute number of reduced confirmed cases is indeed large. 
Figure~\ref{simulation_hubei} (right) displays the numbers of cumulative confirmed cases on day 30 for all 17 cities, when parameters take default values. Our method does not uniformly reduce the confirmed cases for all cities. Instead, our method puts major efforts on reducing the confirmed cases of Huhan and Huanggang, which results in a significant reduction of overall confirmed cases in all cities. For cities such as Xiaogan and Jingzhou, our method yields a larger number of confirmed cases, compared with allocation by infection rate. This is different from the situation for Massachusetts, where the benefit our method is universal for all counties. Figure~\ref{allocation_hubei_1} shows the nonzero entries of $\mathbf{a}(t)$ for $t\in \{3,6,\ldots,27,30\}$. Interestingly, our method puts all testing resources on Wuhan and Ezhou. 
As we have mentioned, Wuhan plays a dominating role in the pandemic progression. It is not surprising that Wuhan is always selected by our method. 
Ezhou is a comparably small city, but it adjoins three big cities, Wuhan, Huanggang and Huangshi, all of which have a large number of confirmed cases. Ezhou also has the largest degree in the commute network. This explains why Ezhou is always selected by our method. We note that this solution is for a daily budget of M = 200K. This is a very limited budget, given that Hubei has a population of 58.9 million. As a result, our method tends to put all testing resources on Wuhan and its close neighbor Ezhou. In the supplemental material, we plot the allocation vectors for M = 500K. The results are more similar to those for Massachusetts.

\subsection{Parameter Estimation} \label{subsec:simu-estimate}

\begin{figure}[tb!]
\centering 
\includegraphics[width=0.9\linewidth]{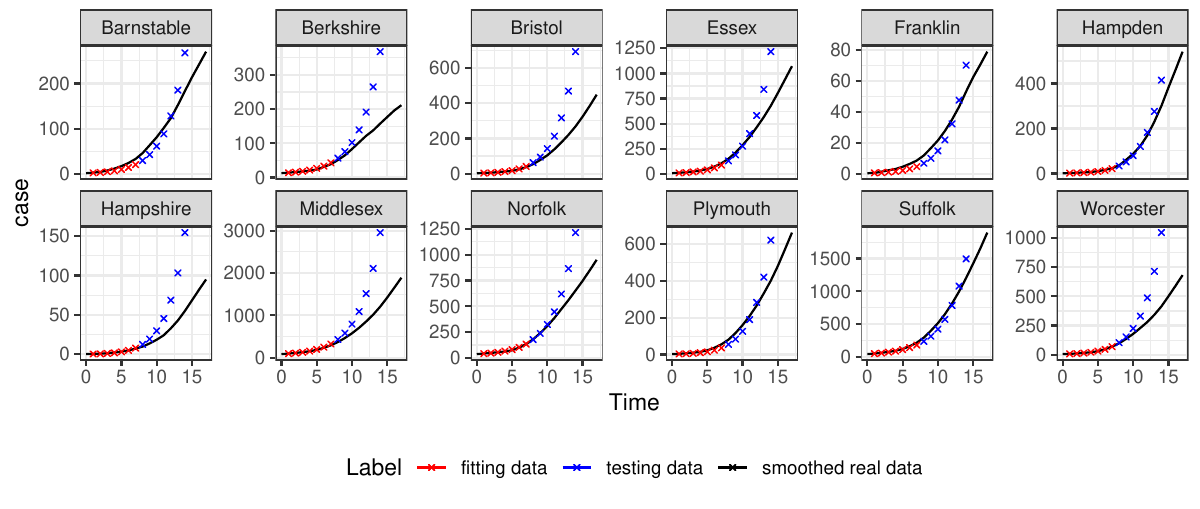}
\caption{Parameter estimation for Massachusetts. The black lines are the smoothed real data of cumulative confirmed cases during March 15--April 1, 2021. The red crosses are the fitting values for March 16-22, 2021, and the blue crosses are the predictions for March 23-29, 2021.} \label{para_est}
\end{figure}

Our screening strategy requires the input of $(\alpha,\beta,\gamma,\lambda, M)$, where $M$ is known and $(\beta,\gamma)$ are obtained from clinical data. We only need to estimate $(\alpha,\lambda)$ from the historical data of confirmed cases. In Section~\ref{subsec:estimation}, we provide two approaches for estimating $(\hat{\alpha},\hat{\lambda})$. We only investigate the second estimator \eqref{Estimate-(alpha+lambda)-3}. The results for the first estimator \eqref{Estimate-(alpha+lambda)-2} are in the supplemental material. 
We fix the time period of March 15 to April 1, 2021 and use daily new confirmed cases of counties in Massachusetts. Our estimator gives $\hat{\alpha} = 0.668$ and $\hat{\lambda} = 55.56 \text{mi}$. 
Since there is no ground truth, we instead evaluate the performance of predicting future confirmed cases. We use $(\hat{\alpha},\hat{\lambda})$ and model \eqref{Mod-ODE-noScr} to forecast the numbers of confirmed cases during March 23-29, 2021 and compare them with the actual reported numbers. The results are in Figure~\ref{para_est}. It is seen that the predictions fit the real data well for Franklin, Hampden and Suffolk. For the other counties, the predicted numbers are higher than the actual reported numbers. 

The results are not surprising. Our model in Section~\ref{subsec:estimation} assumes there is no intervention. However, in reality, Massachusetts took actions such as sending students home and requesting quarantines of travelers  to help slow down the spread of disease. This explains why the predicted numbers are higher than the actual numbers. It also suggests that the administrator should frequently update the estimates using a moving window, because $(\alpha,\lambda)$ can vary with time, as a response to the change of disease control measures.

\section{Extension to the allocation of vaccines} \label{sec:Vaccine}

Vaccination is another way to control disease spread. When the daily capacity of vaccination is limited, we are interested in optimizing the allocation of vaccines. Existing literatures proposed vaccine allocation strategies based on the contact tracing network of individuals \citep{ZhangYao2015} or socioeconomic features such as age and whether being a healthcare worker \citep{Talbot2005}.
We instead consider the vaccine allocation among counties. Similarly as before, we assume that the administrator has historical data of $\{\mathbf{C}^{new}(t)\}_{1\leq t\leq t_0-1}$ and needs to decide a vaccination strategy $\mathbf{v}(t)=(v_1(t),v_2(t),\ldots,v_n(t))^{\top}$ for every $t_0\leq t\leq T$, where $v_i(t)$ is the {\it vaccination rate} allocated to county $i$ at time $t$. We first extend the model in \eqref{Mod-InfectRate}-\eqref{Mod-ODE}. The main difference between screening and vaccination is that the screening intervention moves individuals from compartment H to compartment C, while the vaccination intervention moves individuals from compartment S to compartment R. The modified model is
\begin{equation} \label{Mod-ODE-Vac}
    \left\{ \begin{split}
\Delta s_i(t) &= -[p_i(t)+v_i(t)] s_i(t) \\
\Delta h_i(t) &=p_i(t) s_i(t) - [\beta+\gamma ] h_i(t) \\
\Delta c_i(t) &= \beta h_i(t) - \gamma c_i(t)\\
\Delta r_i(t) &= \gamma [h_i(t)+c_i(t)] + v_i(t) s_i(t)
\end{split} \right., \  \mbox{for every }1\leq i\leq n,
\end{equation}
where $p_i(t)$ is the same as in \eqref{Mod-InfectRate}. Similarly as in \eqref{Proxy-h_i(t)}, we can derive a proxy inequality (`$\leq$' holds for each entry):
\begin{equation} \label{Proxy-h_i(t)-vac}
\mbox{$h_i(t+1) \leq  \mathbf{e}_i^{\top}\mathbf{U}_{vac}(t) \mathbf{h}(t)$}, \ \mbox{where}\quad \mbox{$\mathbf{U}_{vac}(t):=(1- \beta -\gamma)\mathbf{I}_n - \sum_{\tau = t_0}^{t}\mathrm{diag}(\mathbf{v}(\tau)) +\alpha \mathbf{W}$}.  
\end{equation}
Suppose the total vaccination budget is $M$. We compute the allocation vector $\mathbf{v}(t_0), \mathbf{v}(t_0+1),\ldots,\mathbf{v}(T)$ from solving the optimization: 
\begin{equation} \label{Obj-vac}
\begin{gathered}
\mbox{$\min_{\mathbf{v}(t_0:T)} F_{vac}^*(\mathbf{a}(t_0:T)):= \mathbf{N}^{\top}\biggl(\sum_{t=t_0}^T \bigl[ \prod_{m=t_0-1}^{t-1}\mathbf{U}_{vac}(m)\bigr]\biggr)\mathbf{h}_0$}, \\ \mbox{subject to:}\   \mbox{$\mathbf{N}^{\top}\mathbf{v}(t) \leq M$}, \  \mbox{for every $t_0\leq t\leq T$}. 
\end{gathered}
\end{equation}
We can similarly develop an Frank-Wolfe algorithm to solve it. Comparing $\mathbf{U}_{vac}(t)$ with $\mathbf{U}(t)$ in \eqref{Proxy-h_i(t)}, we see the difference between the vaccination strategy and the screening strategy: The past screening allocations $\{\mathbf{a}(\tau)\}_{\tau<t}$ will not affect $\mathbf{U}(t)$, but the past vaccination allocations $\{\mathbf{v}(\tau)\}_{\tau<t}$ will affect $\mathbf{U}_{vac}(t)$. 
The reason is that a susceptible individual can get multiple rounds of screening but only one round of vaccination; consequently, we cannot neglect the past vaccination information, unlike in the situation of screening.

\section{Discussion}\label{sec:Discuss}
We consider the allocation of testing resources on a commute network of counties. We model the progression of pandemic using a 4-compartment network-SIR model, that simultaneously models the effects of  network structure and screening intervention. We propose algorithms for estimating model parameters and computing the optimal allocation strategy. We evaluate their performances on real data for Massachusetts and Hubei.

There are future directions for extending our work. First, we assume that the infection rate $\alpha$ and network effect parameter $\lambda$ are time-invariant. In real life, they may change with the disease control measures such as travel restrictions and quarantine policies \citep{yan2021better}. It is interesting to extend our model to time-varying $(\alpha,\lambda)$. In fact, in our current method, by using a moving window for parameter estimation and letting $T=t_0+1$ for strategy computation, it does support a timely and frequent update of testing resource allocation. 
Second, our screening strategy only considers `efficiency' but not `fairness'. As a result, the strategy computed by our method tends to put testing resources on a small number of counties each day. 
We can modify our method to encourage `fairness'. One possibility is adding a penalty $\delta \sum_{t=t_0}^T\sum_{i=1}^n[a_i(t)-\bar{a}(t)]^2$ to the objective \eqref{Obj-2}, where $\bar{a}(t)=n^{-1}\sum_{i=1}^n a_i(t)$ and $\delta>0$ controls the trade-off between `efficiency' and `fairness'. Our optimization algorithm can be easily extended to accommodate this penalty. Third, our disease model uses no demographic variables of counties.  We may borrow the deep neural network approach in \cite{tang2021interplay} to incorporate county-wise feature vectors into our network-SIR model.

\vspace*{-.5cm}

\bibliography{COVID}

\begin{thebibliography}{}

\bibitem[Baunez et~al., 2020]{Baunez2020}
Baunez, C., Degoulet, M., Luchini, S., Pintus, P., and Teschl, M. (2020).
\newblock {Sub-national allocation of COVID-19 tests: An efficiency criterion
  with an application to Italian regions}.
\newblock {\em Available at SSRN 3576161}.

\bibitem[Becker, 1976]{Becker1976}
Becker, N. (1976).
\newblock Estimation for an epidemic model.
\newblock {\em Biometrics}, 32(4):769--777.

\bibitem[Beigel et~al., 2020]{Beigel2020}
Beigel, J.~H., Tomashek, K.~M., Dodd, L.~E., Mehta, A.~K., Zingman, B.~S.,
  Kalil, A.~C., Hohmann, E., Chu, H.~Y., Luetkemeyer, A., Kline, S., Lopez~de
  Castilla, D., Finberg, R.~W., Dierberg, K., Tapson, V., Hsieh, L., Patterson,
  T.~F., Paredes, R., Sweeney, D.~A., Short, W.~R., Touloumi, G., Lye, D.~C.,
  Ohmagari, N., Oh, M.-d., Ruiz-Palacios, G.~M., Benfield, T., Fatkenheuer, G.,
  Kortepeter, M.~G., Atmar, R.~L., Creech, C.~B., Lundgren, J., Babiker, A.~G.,
  Pett, S., Neaton, J.~D., Burgess, T.~H., Bonnett, T., Green, M., Makowski,
  M., Osinusi, A., Nayak, S., and Lane, H.~C. (2020).
\newblock {Remdesivir for the treatment of Covid-19 - final report}.
\newblock {\em New England Journal of Medicine}, 383(19):1813--1826.

\bibitem[Buhat et~al., 2021]{buhat2020optimal}
Buhat, C. A.~H., Duero, J. C.~C., Felix, E. F.~O., Rabajante, J.~F., and
  Mamplata, J.~B. (2021).
\newblock {Optimal allocation of COVID-19 test kits among accredited testing
  centers in the Philippines}.
\newblock {\em Journal of healthcare informatics research}, 5(1):54--69.

\bibitem[Byambasuren et~al., 2020]{Byambasuren2020}
Byambasuren, O., Cardona, M., Bell, K., Clark, J., McLaws, M.-L., and Glasziou,
  P. (2020).
\newblock {Estimating the extent of asymptomatic COVID-19 and its potential for
  community transmission: systematic review and meta-analysis}.
\newblock {\em Official Journal of the Association of Medical Microbiology and
  Infectious Disease Canada}, 5(4):223--234.

\bibitem[Hao et~al., 2020]{hao2020reconstruction}
Hao, X., Cheng, S., Wu, D., Wu, T., Lin, X., and Wang, C. (2020).
\newblock {Reconstruction of the full transmission dynamics of COVID-19 in
  Wuhan}.
\newblock {\em Nature}, 584(7821):420--424.

\bibitem[Li et~al., 2020]{Li2020}
Li, Q., Guan, X., Wu, P., Wang, X., Zhou, L., Tong, Y., Ren, R., Leung, K.~S.,
  Lau, E.~H., Wong, J.~Y., Xing, X., Xiang, N., Wu, Y., Li, C., Chen, Q., Li,
  D., Liu, T., Zhao, J., Liu, M., Tu, W., Chen, C., Jin, L., Yang, R., Wang,
  Q., Zhou, S., Wang, R., Liu, H., Luo, Y., Liu, Y., Shao, G., Li, H., Tao, Z.,
  Yang, Y., Deng, Z., Liu, B., Ma, Z., Zhang, Y., Shi, G., Lam, T.~T., Wu,
  J.~T., Gao, G.~F., Cowling, B.~J., Yang, B., Leung, G.~M., and Feng, Z.
  (2020).
\newblock {Early transmission dynamics in Wuhan, China, of novel
  coronavirus-infected pneumonia}.
\newblock {\em New England Journal of Medicine}, 382(13):1199--1207.

\bibitem[Ou et~al., 2020]{ou2020and}
Ou, H.-C., Sinha, A., Suen, S.-C., Perrault, A., Raval, A., and Tambe, M.
  (2020).
\newblock Who and when to screen: Multi-round active screening for network
  recurrent infectious diseases under uncertainty.(2020).
\newblock In {\em {Proceedings of 19th International Conference on Autonomous
  Agents and Multiagent Systems (AAMAS), Auckland New Zealand}}, pages 9--13.

\bibitem[Talbot et~al., 2005]{Talbot2005}
Talbot, T.~R., Bradley, S.~F., Cosgrove, S.~E., Ruef, C., Siegel, J.~D., and
  Weber, D.~J. (2005).
\newblock Influenza vaccination of healthcare workers and vaccine allocation
  for healthcare workers during vaccine shortages.
\newblock {\em Infection Control \& Hospital Epidemiology}, 26(11):882--890.

\bibitem[Tang et~al., 2021]{tang2021interplay}
Tang, F., Feng, Y., Chiheb, H., and Fan, J. (2021).
\newblock {The interplay of demographic variables and social distancing scores
  in deep prediction of US COVID-19 cases}.
\newblock {\em Journal of the American Statistical Association},
  116(534):492--506.

\bibitem[Wang et~al., 2020]{wang2020epidemiological}
Wang, L., Zhou, Y., He, J., Zhu, B., Wang, F., Tang, L., Kleinsasser, M.,
  Barker, D., Eisenberg, M.~C., and Song, P.~X. (2020).
\newblock {An epidemiological forecast model and software assessing
  interventions on the COVID-19 epidemic in China}.
\newblock {\em Journal of Data Science}, 18(3):409--432.

\bibitem[Wang et~al., 2003]{wang2003epidemic}
Wang, Y., Chakrabarti, D., Wang, C., and Faloutsos, C. (2003).
\newblock Epidemic spreading in real networks: An eigenvalue viewpoint.
\newblock In {\em 22nd International Symposium on Reliable Distributed Systems,
  2003. Proceedings.}, pages 25--34. IEEE.

\bibitem[Yan et~al., 2021]{yan2021better}
Yan, H., Zhu, Y., Gu, J., Huang, Y., Sun, H., Zhang, X., Wang, Y., Qiu, Y., and
  Chen, S.~X. (2021).
\newblock {Better strategies for containing COVID-19 pandemic: a study of 25
  countries via a vSIADR model}.
\newblock {\em Proceedings of the Royal Society A}, 477(2248):20200440.

\bibitem[Yin and B{\"u}y{\"u}ktahtak{\i}n, 2021]{Yin2021}
Yin, X. and B{\"u}y{\"u}ktahtak{\i}n, {\.I}.~E. (2021).
\newblock A multi-stage stochastic programming approach to epidemic resource
  allocation with equity considerations.
\newblock {\em Health Care Management Science}, pages 1--26.

\bibitem[Zhang and Prakash, 2015]{ZhangYao2015}
Zhang, Y. and Prakash, B.~A. (2015).
\newblock Data-aware vaccine allocation over large networks.
\newblock {\em ACM Transactions on Knowledge Discovery from Data (TKDD)},
  10(2):1--32.

\end{thebibliography}



\section*{Supplementary material}

# Supplementary Material to Allocation of COVID-19 Testing Budget on a Commute Network of Counties

Yaxuan Huang[1], Zheng Tracy Ke*[2], and Jiashun Jin[3]

[1]Department of Statistics, University of California, Berkeley, California, USA
[2]Department of Statistics, Harvard University, Massachusetts, USA
[3]Department of Statistics, Carnegie Mellon University, Pennsylvania, USA

This is the supplementary to paper *Allocation of COVID-19 Testing Budget on a Commute Network of Counties*. In Section A and B, we provide additional information for the synthetic experiments on the commute network of Massachusetts and Hubei. And in Section C, we provide some additional information for the parameter estimation.

## A  Additional information for Massachusetts

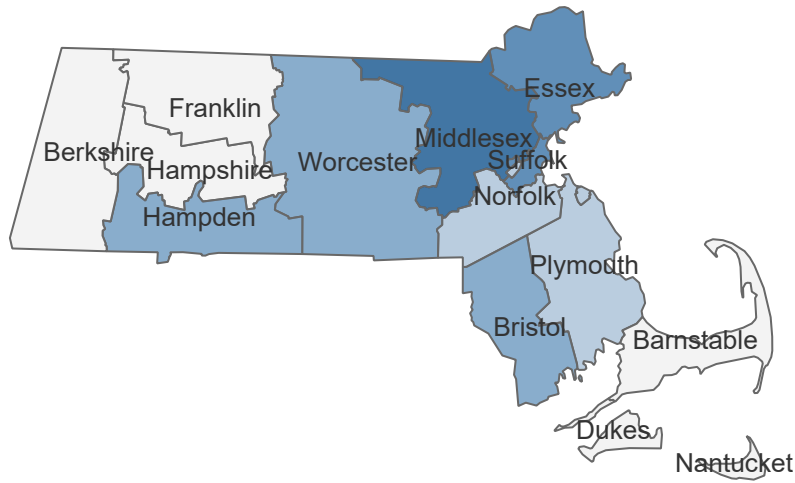

**Figure S1:** COVID-19 map of Massachusettes county. The color denotes the culmulative confirmed cases at September 30th, 2020. Darker blue means larger number of cases.

In Section 3.1 of the main paper, we provide our experiment results on the commute network of Massachusettes network and Figure 2(left) show the network structure constructed based on traffic distance. Here we provide the map of Massachusetts in Figure S1. This is download from the COVID-19 Interactive Data Dashboard from Mass.gov. It shows the culmulative confirmed case numbers of counties in Massachusettes at September 30th, 2021. This is different from our

experiment settings. In our synthetic analysis, we use the confirmed case numbers of March 19th and 20th, 2020, the beginning stage of COVID-19 pandemic, as the start point of our simulation.

Based on the commute network, we conduct simulations under different strategies and compare the strategy proposed by our screening strategy with the allocation strategy based infection rates. We present 10 days allocation in Figure 4 in the main paper, and in Figure S2, we present the full plot under our default seting for Massachusettes, $\lambda = 100$, $\beta = 0.16$, $\gamma = 1/15$, $\alpha = 0.3$, M = 100K.

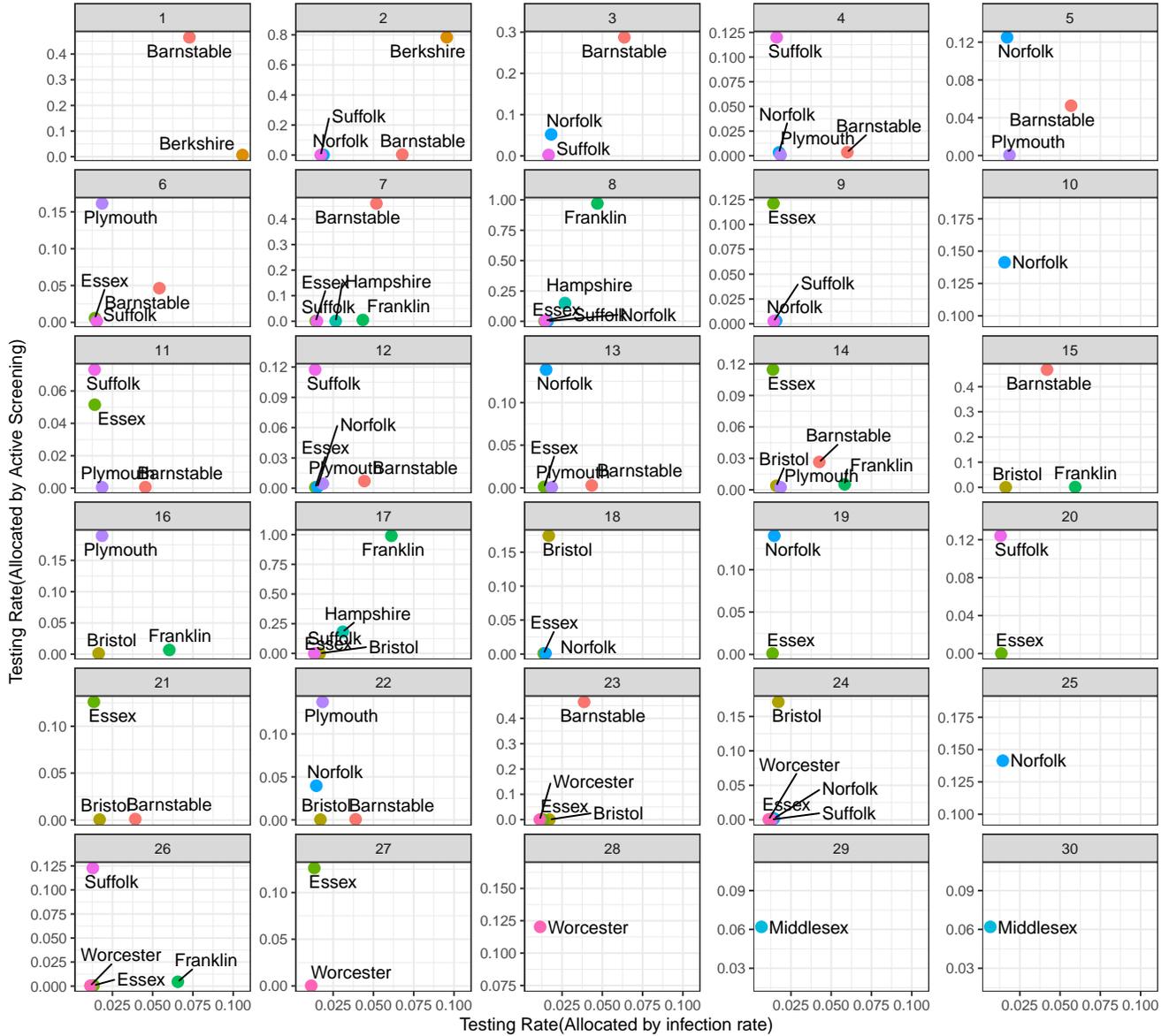

**Figure S2:** Comparison of the allocated testing rate by our method versus the infection rate in Massachusetts. We only plot counties with a non-zero testing rate in our method. The parameters are $\lambda = 100$, $\beta = 0.16$, $\gamma = 1/15$, $\alpha = 0.3$, M = 100K.

2

# B  Additional information for Hubei

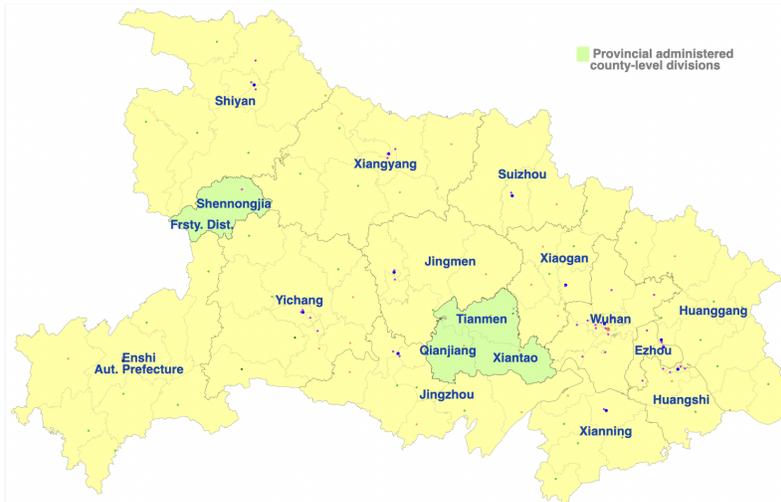

**Figure S3:** City map of Hubei province (Download from Wikipedia.).

Similarly, we also conduct simulation on commute network of Hubei. The network structure in Figure 2(right) is constructed based on the map of Hubei shown in Figure S1. This is download from Wikipedia. In our synthetic analysis of Hubei, we use the confirmed case numbers of Febrary 1st and 2nd, 2020, the beginning stage of COVID-19 pandemic, as the start point of our simulation.

Based on the commute network, we conduct simulations under different strategies and compare the strategy proposed by our screening strategy with the allocation strategy based infection rates. We present 10 days allocation in Figure 4 in the main paper, and in Figure S4, we present the full plot under our default seting for Massachusettes, $\lambda = 50$, $\beta = 0.16$, $\gamma = 1/15$, $\alpha = 0.3$, M = 200K. As we mentioned in the Section 3.1 of the main paper, M = 200K is not enough for such big population in Hubei province. And in this case, our algorithm will choose an relatively extreme strategy by concentrating all resources on Wuhan and Ezhou. If the budget increase, say M = 500K, the algorithm will provide similar results as our discussion for Massachusetts. Figure S4 presents allocation strategies on this setting ($\lambda = 50$, $\beta = 0.16$, $\gamma = 1/15$, $\alpha = 0.3$, M = 500K). In this case,besides Wuhan and Ezhou, the algorithm will also allocate testing resources to cities like Huangshi, Suizhou and Xiaogan, which are also key nodes for the commute network of Hubei.

# C  Additional information for parameter estimation

In the Section 2.3 of the main paper, we propose two methods for estimating $\lambda$ and $\alpha$. The estimation results on real data of the second method are discussed in Section 3.2 in the main paper. Here we show the estimation results of the first method. Using the same data preriod as main paper, from March 15, 2021 to April 1, 2021, the optimal parameters given by the first method is $\hat{\alpha} = 0.50$ and $\hat{\lambda} = 19.61 mi$. This infection rate is smaller than the results given by the second method. Therefore, in fitting and prediction, it tends to underestimate the number of confirmed cases as shown in Figure S6.



\end{document}